\documentclass[11pt,twoside]{article}

\usepackage{asp2006}
\usepackage{graphicx}
\usepackage{psfig}
\usepackage{lscape}

\newcommand{\be}{\begin{equation}}
\newcommand{\ee}{\end{equation}}
\newcommand{\bea}{\begin{eqnarray}}
\newcommand{\eea}{\end{eqnarray}}
\newcommand{\ba}{\begin{array}}
\newcommand{\ea}{\end{array}}

\newcommand{\lsim}{\buildrel < \over {_\sim}}
\newcommand{\gsim}{\buildrel > \over {_\sim}}

\begin{document}
\title{Minimal Electroweak Scale Cosmology and the LHC}   
\author{Michael Ramsey-Musolf}   
\affil{Department of Physics\\ University of Wisconsin-Madison\\ Madison, WI 53706}    

\begin{abstract} 
I discuss simple extensions of the Standard Model scalar sector that can help solve the origin of matter problem and that can be probed in their cosmologically relevant parameter space at the LHC.
\end{abstract}

\section{Introduction}
\label{sec:intro}
One of the particle physics community's hopes for the Large Hadron Collider (LHC) is that its discoveries will help solve some of the open problems in cosmology. In this talk, I will focus on the yet unexplained origin of matter (both visible and dark) and ask the following questions: To what extent can minimal extensions of the Standard Model (SM) provide a particle physics explanation for the origin of matter? And to what extent might one these models be discovered and probed at the LHC?

In addition to explaining the origin of matter, one also hopes that LHC discoveries will address other problems, such as the gauge hierarchy problem, unification, the tension between the direct search lower bound on the Standard Model Higgs boson mass and electroweak precision observables (EWPO), the origin of neutrino masses, and so on. I like to classify various approaches to solving these problems in two broad categories: non-minimal solutions such as supersymmetry that can have some well-defined U.V. completion, and minimal solutions that will be the focus of this discussion. 


Supersymmetry in the guise of the MSSM is a highly-favored \lq\lq minimal" non-minmal scenario, in part because of its potential to solve nearly all of the open problems mentioned, not to mention providing unification, radiative electroweak symmetry-breaking, and a natural embedding in string theory. Its benefits, however, come with a cost: the $\mu$-problem, the SUSY flavor and CP-problems, the large number of new parameters that beg for a fundamental explanation, {\em etc} -- all of which have kept a large number of theorists gainfully employed over the years. From the standpoint of the origin of matter, the MSSM is also tightly constrained: the gauginos should be lighter that one TeV to avoid Boltzmann suppression during the era of electroweak symmetry-breaking; the conventional neutralino dark matter candidate should be bino-Higgsino like in order to achieve the observed relic density; the Standard Model-like Higgs boson and right-handed stop should both be lighter than about 125 GeV; the gaugino mass parameter $M_1$ should be close to $\mu$ in magnitude; and the CP-violating phases associated with the electroweak gauginos should be non-universal. 

Given these constraints as well as the new problems SUSY introduces, it may be worth our while to consider the simplest extensions of the SM which have the potential to help solve as many of the observational problems as the MSSM though not necessarily those of theoretical interest, such as the hierarchy problem and unification. Here, I will take the point of view that such minimal extensions may be low-energy remnants of a more complete theory that also addresses the theoretical problems and that by studying them, we may gain some insight into the generic features of realistic models that address yet unexplained observational facts without being encumbered by their model-dependent complications. To that end, I will focus on simple extensions of the scalar sector of the SM that entail introducing one, two, or three new fields to the SM: the \lq\lq xSM" involving one additional real, scalar gauge singlet; the \lq\lq cxSM" that adds a complex singlet; and the \lq\lq $\Sigma$SM" obtained with the SM plus a real SU(2)$_L$ triplet. 

\section{Visible Matter: Electroweak Baryogenesis }
\label{sec:baryo}

Since the theory and phenomenology of dark matter is likely quite familiar to this audience, I will briefly review the requirements for a particle physics explanation of the origin of visible matter, or baryogenesis. Over four decades ago, Sakharov identified the necessary ingredients for successful baryogenesis\citep{Sakharov:1967dj}: violation of baryon number (B); violation of both C and CP-symmetry; and departure from equilibrium dynamics, assuming that CPT is a symmetry of nature. In gauge theories, B-violation occurs through anomalous processes called sphalerons. C- and CP-violation are needed for the generation of charge asymmetries in the early universe that drive the sphalerons into making baryons; and a departure from equilibrium is needed to prevent a washout of net baryon number by inverse processes. Although the SM in principle contains all three ingredients, the effects of CKM CP-violation are too feeble to lead to sufficiently large charge asymmetries. Moreover, the LEP lower bound $m_H < 114.4$ GeV precludes a cosmic phase transition associated with electroweak symmetry-breaking (EWSB) in the SM universe. Consequently, even if the CP-violating asymmetries had been sufficiently large, all of their effects would have been erased by sphalerons. 

Experimentally, searches for permanent electric dipole moments and CP-violation in neutrino interactions with long baseline oscillation studies hope to uncover the CP-violation needed for successful baryogenesis.
Searches for new scalars at the LHC may provide clues about the physics needed for a phase transition during the EWSB era. Since the focus of this talk is on the LHC, let me explain how the latter works.

The dynamics of symmetry-breaking at finite temperature are governed by the finite-temperature effective potential, $V_\mathrm{eff}(\varphi, T)$, where $\varphi$ denotes the classical field. In the SM, $\varphi$ is just the vacuum expectation value of the neutral Higgs field. In this case, the ring-improved, SM one-loop effective potential has the simple form in the high temperature limit [for a pedagogical review, See Ref.~ \citep{Quiros:1999jp}]:
\be
\label{eq:veffsm}
V_\mathrm{eff}^\mathrm{SM}(\varphi, T) = 2D(T^2-T_0)^2 \varphi^2 - 2\sqrt{2} E T\varphi^3 + \lambda \varphi^4 +\cdots\ \ \ ,
\ee
where $\lambda$ is the Higgs quartic self-coupling; $D$, $E$, and $T_0^2$ are calculable in perturbation theory in terms of $\lambda$ and the gauge couplings so long as $\lambda$ is not too large; and the \lq\lq $+\cdots$ denote small corrections in the high-$T$ limit. Note that the cubic term only appears at $T\not=0$; its coefficient arises at one-loop order from loops involving the transverse components of the $W^\pm$ and $Z$ bosons.  At high temperatures, the minimum of $V_\mathrm{eff}^\mathrm{SM}(\varphi, T)$ is at the origin ($\varphi=0$). As the universe cools, a secondary minimum for $\varphi\not=0$ develops, and at some critical temperature $T_C$ the two minima become degenerate. Slightly below this temperature, bubble nucleation commences as the EWPT proceeds. 

Assuming new CP-violating interactions produced sufficiently large asymmetries to drive the sphalerons into making baryons, the phase transition must quench the sphalerons inside the expanding bubbles of broken electroweak symmetry in order to preserve the baryon asymmetry. In the standard treatments, this requirement implies that the value $\varphi$ at the nucleation temperature $T_N$ satisfy:
\be
\label{eq:phitc}
\frac{\varphi(T_N)}{T_N} \gsim \frac{\varphi(T_C)}{T_C}\gsim 1 \ \ \ .
\ee
From Eq.~(\ref{eq:veffsm}) one can show that Eq.~(\ref{eq:phitc}) implies
\be
\label{eq:ewptreq}
\frac{2E}{\lambda} \gsim 1 \ \ \ .
\ee
In short, the finite temperature-induced cubic term in $V_\mathrm{eff}^\mathrm{SM}(\varphi, T) $ must be sufficiently large compared to the quartic self coupling in order to prevent washout. However, one can also express $\lambda$ in terms of the zero-temperature Higgs vev $v=246$ GeV and Higgs mass: $m_H^2=2\lambda v^2$, implying that
\be
\label{eq:ewpt1}
4E\left(\frac{v}{m_H}\right)^2 \gsim 1 \ \ \ .
\ee
In perturbation theory, one has $4E=0.038$ so that the Higgs must be lighter than about 45 GeV to satisfy the washout criterion. Nonperturbative computations suggest that this bound is somewhat larger, but it is still below the LEP lower bound on $m_H$.

A strong, first order EWPT satisfying the criterion of Eq.~(\ref{eq:phitc}) can arise in extensions of the SM having additional scalars. In the MSSM, for example, top squark loops may enhance the value of $E$ if the scalar partner of the right-handed top, ${\tilde t}_R$, is sufficiently light. Stops have the largest effect since they have $\mathcal{O}(1)$ couplings to the Higgses and since, for each chiral multiplet, they introduce $2 N_C$ loops (the factor of two arises for a complex scalar). Thus, from perturbation theory, one might expect that the upper bound on the lightest SM-like Higgs mass can be about a factor of two to three larger than for the SM, a situation born out by careful analyses [see Ref.~\citep{Carena:2008vj} and references therein]. 

The left-handed stops must be heavy in order to give sufficiently large contributions to the lightest Higgs mass, but the right-handed stops will be effective if their mass is less than about 125 GeV. In order for ${\tilde t}_R$ to lie in this range, one must make a special choice of the soft SUSY-breaking mass parameter associated with its mass so that it is lighter than the top quark. Moreover, one has to be careful that the RH stop mass parameter does not lead to dangerous color-breaking minima of the potential.  Theoretically, this scenario may not be too attractive. Phenomenologically, however, it has motivated considerable interest in searching for light stops at the Tevatron and LHC. 

An alternate possibility is to expand the scalar sector associated with symmetry breaking and search for new interactions that will enable Eq.~(\ref{eq:phitc}) to be satisfied while also respecting the LEP bound on $m_H$. In what follows, I will discuss how new interactions can do so by either effectively increasing the numerator of Eq.~(\ref{eq:ewptreq}) or decreasing the denominator. Moreover, these new interactions can have observable consequences for the LHC and in the case of the cxSM, may also allow one to explain the dark matter relic density. 

\section{Extended Scalar Sectors and the EWPT}
\label{sec:scalar}

Let me illustrate the basic ideas with the simplest extension: the xSM that involves adding a single real scalar $S$ to the SM scalar sector. There exist two gauge invariant, renormalizable operators involving $S$ and the SM Higgs doublet $H$, that one can include in the scalar potential\citep{Profumo:2007wc,O'Connell:2006wi,Barger:2007im}:
\be
\label{eq:vhs}
V_{HS} = \frac{a_1}{2} H^\dag H S + \frac{a_2}{2} H^\dag H S^2 \ \ \ .
\ee
For simplicity, I do not include the other terms in the potential that involve only $H$ or $S$ alone. In general both $H$ and $S$ may get vevs. Let $\langle H\rangle = v/\sqrt{2}$, $\langle S \rangle  = x$, and their zero-temperature values be $v_0/\sqrt{2}$ and $x_0$, respectively. Note that when $a_1\not=0$, one must have $x\not=0$. In this case, $S$ is not stable, and the model includes no dark matter candidate. It has been shown that for scenarios wherein $x=0$, strengthening the EWPT for successful baryogenesis requires of order ten real scalars that contribute to $E$ only through loop effects. Here, I will consider scenarios wherein $x\not=0$. 

\begin{figure}
\begin{center}
\includegraphics[scale=0.25]{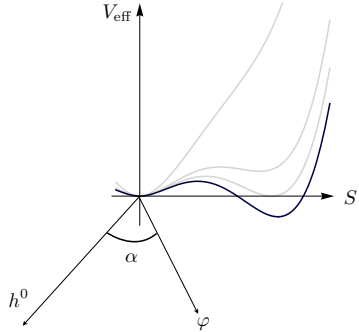}
\end{center}
\caption{Finite temperature effective potential in the xSM as a function of the classical SU(2) ($h^0$) and singlet ($S$) fields. Grey and black curves show schematic evolution from high to low temperature (Courtesy H. Patel). }
\label{fig:veff}
\end{figure}

To analyze the pattern of symmetry breaking at finite temperature, one must consider the behavior of the effective potential in the two dimensional space of $S$ and $h^0$ (the neutral component of the doublet), as illustrated in Fig. \ref{fig:veff}. It is convenient to work with a new set of fields $\varphi$ and $\alpha$, corresponding to cylindrical co-ordinate as shown. For purposes of quenching the SU(2$)_L$ sphalerons, what matters is the projection of $\varphi$ along the $h^0$ axis at $T\sim 100$ GeV. From the modified potential, one finds that the criteria that this projection be larger than $T_C$ is now\citep{Profumo:2007wc}
\be
\label{eq:ewpt2}
\sqrt{2}\cos\alpha_C \left(\frac{\varepsilon -e/T_C}{2{\bar\lambda}}\right)+\cdots \gsim 1\ \ \ ,
\ee
where $\alpha_C$ is the value of the field $\alpha$ at $T_C$; where $\varepsilon  =  2\sqrt{2} E \cos\alpha_C^3$ is the SM loop-induced cubic coefficient;  where
\be
\label{eq:ewptparam}
e  =  \frac{a_1}{2}\cos^2\alpha_C \sin\alpha_C +\cdots, \qquad 
{\bar\lambda} \approx  \lambda \cos\alpha_C^4 +\frac{a_2}{2} \cos^2\alpha_C\sin^2\alpha_C +\cdots;\ \ \ 
\ee
and where the \lq\lq $+\cdots$" denote either small corrections at high $T$ or terms involving the singlet potential that are not important for this discussion. In arriving at Eq.~(\ref{eq:ewpt2}), we have assumed that as the universe cools, the potential breaks directly to a minimum in which both $v$ and $x$ are non-zero. Under some conditions, it is possible that one first breaks to a minimum in which $v=0$ and $x\not=0$ followed by a transition to the broken singlet and electroweak minimum. The EWPT is even stronger in the latter case, which I do not discuss here.

From Eqs.~(\ref{eq:ewpt2},\ref{eq:ewptparam}) one can observe two ways in which the new interactions can strengthen the EWPT: (1) the $e$-term that arises from the new cubic interaction in Eq.~(\ref{eq:vhs}) is both negative and sufficiently large in magnitude, thereby increasing the numerator of Eq.~(\ref{eq:ewpt2}) over that in Eq.~(\ref{eq:ewptreq}); (2) the second term in ${\bar\lambda}$ is negative and sufficiently large in magnitude, thereby decreasing the denominator in Eq.~(\ref{eq:ewpt2}) from that in Eq.~(\ref{eq:ewptreq}). Numerical simulations using the full finite-temperature effective potential confirm the expectations based in this high-$T$ analysis\citep{Profumo:2007wc}. 

\section{LHC Phenomenology \& Other Minimal Scenarios}
\label{eq:lhc}
Searches for new scalars at the LHC can be used to probe the new interactions of the form given in Eq.~(\ref{eq:vhs}) through two basic effects: (1) mixing of the $h^0$ and $S$ at $T=0$ to form new neutral mass eigenstates, $h_1$ and $h_2$, thereby reducing the production strength of either compared to that of a SM Higgs boson; (2) the possibility that either of these scalars may decay to two of the others, thereby leading to changes in branching ratios from their SM expectations. Both effects can be characterized by the \lq\lq signal reduction factor" $\xi_j^2$\citep{Barger:2007im}:
\be
\xi_j^2=V_{1j}^2 \frac{\mathrm{BF}(h_j\to X_\mathrm{SM})}{h_\mathrm{SM}\to X_\mathrm{SM})}\ \ \ ,
\ee
where $V_{1j}^2$ is the SU(2$)_L$ fraction of the state $h_j$ and $X_\mathrm{SM}$ is a given SM final state to which the SM Higgs scalar $h_\mathrm{SM}$ of a given mass should decay. 

To understand how the LHC could probe the parameters related to the EWPT using these two effects, first consider a special case in which $a_1=0$ but $x=0$ (note that this is an illustrative, rather than realistic case because the spontaneous breaking of $Z_2$ symmetry would give rise to unacceptable cosmological domain walls). In this instance, the EWPT can be strengthened only via the effect of the new quartic interaction that enters the denominator of Eq.~(\ref{eq:ewpt2}). From our numerical study\citep{Profumo:2007wc}, we find that those parameter choices yielding a strong first order EWPT that are consistent with both the LEP bound on the SM-like Higgs and constraints from electroweak precision observables (primarily the oblique parameters) also lie in the range where the mass of the SM-like scalar, $m_1$, is greater than twice the mass of the singlet-like scalar, $m_2$. In this region, the decay $h_1\to h_2 h_2$ is kinematically allowed. Moreover, we find that the branching fraction for this decay is typically  $\gsim 50\% $, implying that $\xi_1^2 \lsim 0.5$, implying a rather dramatic change in the expected number of SM-like Higgs scalars in traditional Higgs searches at the LHC. 

In the more general case, when $a_1\not=0$, a strong first order EWPT can occur either via the effect of the new quartic interaction or the cubic interaction proportional to $a_1$. The phenomenological effect of the latter is to generate mixing between $h^0$ and $S$. From our numerical studies in Ref.~\citep{Barger:2007im}, we asked the following questions: Can a mixed doublet-singlet scalar be discovered at the LHC using traditional Higgs boson searches? If so, how well can one determine the value of $\xi_j^2$ and, thereby obtain a handle on the impact of both mixing and changes in the branching ratio as implied by the new interactions of Eq.~(\ref{eq:vhs})? We found that for the xSM, one could, indeed, discover one or the other of the scalars with greater than $5\sigma$ significance with $\xi_j^2$ as low as $\sim  0.5$ with 30 $\mathrm{fb}^{-1}$ integrated luminosity. Moreover, depending on the mass of the discovered scalar, one could even determine the value of $\xi_j^2$ with reasonable uncertainty. In short, with a few years of LHC running, one could determine if minimal new scalar interactions are present and if they have the right strengths to lead to a strong first order EWPT.


While the xSM illustrates the key features of simple scalar sector extensions as they pertain to the EWPT and LHC, it may be more realistic to consider complex singlets or simple representations that carry SU(2$)_L$ quantum numbers. The introduction of a complex singlet $\mathcal{S}$ has the advantage that one can achieve a similar strengthening of the EWPT as in the xSM as well as a viable dark matter candidate\citep{Barger:2008jx}. The most general, renormalizable potential involving $H$ and $\mathcal{S}$ contains three interactions between the two fields: $H^\dag H\mathcal{S}+\mathrm{c.c}$, $H^\dag H\mathcal{S}^2+\mathrm{c.c}$, and $H^\dag H\vert\mathcal{S}\vert^2$. In order to obtain viable dark matter candidate in this model, we first eliminate the cubic interaction as well as all cubic terms in $\mathcal{S}$ by imposing a global U(1) symmetry. Doing so leaves only the quartic interaction involving $\vert\mathcal{S}\vert^2$. For suitable parameter choices, spontaneous symmetry breaking (SSB) that gives a vev to  $\mathcal{S}$ and breaks the global U(1), leading to
\be
\label{eq:ssb}
\delta_2 H^\dag H\vert\mathcal{S}\vert^2\rightarrow \frac{\delta_2}{2} H^\dag H \left[ (x+S)^2+A^2\right]\ \ \ ,
\ee
where $\delta_2$ is the quartic coupling and $S/\sqrt{2}$ and $A/\sqrt{2}$ denote the real and imaginary parts of $\mathcal{S}$ after shifting by its vev $x/\sqrt{2}$, with $A$ being the Goldstone boson of the spontaneously broken U(1). To make the latter a viable dark matter particle, we give it a mass by softly breaking the U(1).  This is our cxSM scenario. 

Note that the term $H^\dag H(x+S)^2$ behaves just like our $a_2$ term in Eq.~(\ref{eq:vhs}) after SSB, while the $H^\dag H A^2$ term leads to the interactions $AA\leftrightarrow hh$, $AA\leftrightarrow h \leftrightarrow f{\bar f}$ {\em etc}. Thus, this interaction can both bring about the necessary reduction in the effective quartic coupling at finite temperature and control the relic density associated with the $A$ via these annihilation processes. From our numerical study\citep{Barger:2008jx}, we observe that its coefficient $\delta_2$ can be chosen to satisfy both requirements of a strong first order EWPT and observed relic density with masses for the real scalars lying above the LEP bound. 

One can also search for this interaction at the LHC either using the study of the signal reduction factor as discussed above or by exploiting another search strategy known as an invisible search. This strategy involves consideration of the production of the real eigenstates $h_j$ in weak boson fusion (WBF): $q{\bar q} \to q{\bar q} W^\ast W^\ast \to q{\bar q} h_j$ and with the $h_j$ decaying invisibly to an $AA$ pair which has no SM interactions aside from those with Higgs bosons. These events are characterized by large missing energy. By observing the azimuthal distribution of the outgoing jets (the quarks that radiated the virtual weak vector bosons), one can distinguish this process from SM  processes in having large missing energy. The strategy is likely to be most successful when the quartic coupling strength $\delta_2$ and scalar masses imply at least a $\sim 50\%$ invisible decay branching fraction, precisely the region in parameter space where one expects the new quartic interaction to give rise to a strong first order EWPT (assuming $\delta_2< 0$).

When one extends the scalar sector with a non-trivial representation of SU(2$)_L$, both the LHC phenomenology and the implications for cosmology can be quite distinct. The smallest representation carrying non-trival SU(2$)_L$ charge is a real triplet, carrying zero hypercharge: $\Sigma\equiv(\Sigma^0, \Sigma^+,\Sigma^-)$. The resulting \lq\lq $\Sigma$SM" \citep{FileviezPerez:2008bj} can give rise to a dark matter candidate -- the $\Sigma^0$ -- if the latter has no vev. In principle, it may yield a first order phase transition as needed for baryogenesis under some scenarios. In practice, constraints from the $\rho$-parameter of electroweak precision phenomenology imply that if the neutral component obtains a vev, it must be very small at $T=0$, making a strong, single-step first order EWPT unlikely. In the dark matter scenario, it appears that one could discover the triplet at the LHC only if it is lighter than about 500 GeV, in which case it could make up no more than about one tenth of the relic density. The presence of gauge interactions, leading to annihilation processes such as 
$\Sigma^0\Sigma^0\to W^+ W^-$, imply too large an annihilation cross section for a light triplet to make up much more of the relic density. On the other hand, the production cross section at the LHC becomes too small to make discovery of a heavier triplet realistic. It may be, however, that the dark sector is more complicated than in a single species scenario. In this case, the real triplet -- well-motivated from a grand unification standpoint -- may be part of a constellation of dark matter particles.

\section{Summary}
\label{sec:sum}

Although I have omitted many details and references due to space limitations, I hope I have conveyed the message that minimal extensions of the SM scalar sector can provide some of the necessary ingredients for explaining the origin of matter and that these scenarios can be probed in their cosmologically relevant parameter space at the LHC. A summary of some of their key features is given in Table 1 below. The discovery of one or more of these minimal scenarios may point us to the complete theory in which they are embedded and which solves both our observational puzzles and unanswered theoretical questions.

\begin{center}
\label{tab1}
\begin{table}[h] 
\begin{tabular}[t]{|c||c|c||c|}
\hline Scenario &  DOF & Cosmology \& EWPO  & LHC  \\
\hline
xSM & 1 & EWPT \& $m_H$/EWPO tension& $\xi_j^2 < 1$   \\
 & & or DM & $\xi_j^2<1$ \& WBF invisible \\
 \hline
 cxSM & 2 & EWPT \& $m_H$/EWPO tension & $\xi_j^2 < 1$   \\
 & & and DM & and WBF invisible \\
 \hline
  $\Sigma$SM & 3 & DM & charged track   \\
 & & potential EWPT  & Excess $\tau\nu\gamma\gamma$ or $b{\bar b}\gamma\gamma$ \\
 \hline
\end{tabular}
\caption{Implications of minimal scalar sector extensions for cosmology, electroweak precision observable (EWPO), and LHC searches.   The xSM and cxSM involve no new gauge interactions, in contrast to the $\Sigma$SM. The second column gives the number of new scalar degrees of freedom beyond those of the SM.}
\end{table}
\end{center}



\acknowledgements 

I would like to thank the SnowPAC organizers for the invitation to speak and my many collaborators in this work: V. Barger, P. Langacker, M. McCaskey, D. O'Connell, H. Patel, P. Fileviez Perez, S. Profumo, G. Shaugnessy, K. Wang, and M. Wise. I also thank H. Patel for producing the figure. This work was supported in part by Department of Energy contract
DE-FG02-08ER41531  and the Wisconsin Alumni Research Foundation.


\end{document}